\documentclass[conference]{IEEEtran}

\usepackage[dvipsnames]{xcolor}
\usepackage{epsfig}
\usepackage{times}
\usepackage{float}
\usepackage{afterpage}
\usepackage{amsmath}
\usepackage{amstext,cite}
\usepackage{amssymb,bm}
\usepackage{latexsym}
\usepackage{color}
\usepackage{graphicx}
\usepackage{amsmath}
\usepackage{amsthm}
\usepackage{graphicx}
\usepackage[center]{caption}
\usepackage{pstricks}
\usepackage{caption}
\usepackage{subcaption} 
\usepackage{booktabs}
\usepackage{multicol}
\usepackage{lipsum}%
\usepackage[T1]{fontenc}
\usepackage{hyperref}
\usepackage{aecompl}
\usepackage{mathrsfs}
\usepackage{enumitem}
\usepackage{sidecap}

\usepackage{tikz}
\usetikzlibrary{shapes.multipart}
\usetikzlibrary{fit}
\usetikzlibrary{shapes.geometric,positioning}
\usetikzlibrary{calc}
\usetikzlibrary{patterns.meta}
\usetikzlibrary{arrows.meta} 

\usepackage{soul}
\usepackage{cancel}

\usepackage{tikz}
\usetikzlibrary{shapes.multipart}
\usetikzlibrary{fit}
\usetikzlibrary{shapes.geometric,positioning}
\usetikzlibrary{calc}
\usetikzlibrary{patterns.meta}
\usepackage{pgfplots}
\usetikzlibrary{plotmarks}
\usetikzlibrary{spy}

\long\def\comment#1{}

\newcommand{\RR}{{\mathbb R}}

\newcommand{\FF}{{\mathbb F}}

\newcommand{\NN}{{\mathbb N}}

\newcommand{\fv}{{\mathbf f}}

\newcommand{\wv}{{\mathbf w}}

\newcommand{\Am}{{\mathbf A}}
\newcommand{\Bm}{{\mathbf B}}

\newcommand{\Fm}{{\mathbf F}}

\newcommand{\Id}{{\mathbf I}}

\newcommand{\Mm}{{\mathbf M}}
\newcommand{\Nm}{{\mathbf N}}

\newcommand{\Um}{{\mathbf U}}
\newcommand{\Wm}{{\mathbf W}}
\newcommand{\Vm}{{\mathbf V}}

\newcommand{\Zm}{{\mathbf Z}}

\newcommand{\Bc}{{\mathcal B}}

\newcommand{\Lc}{{\mathcal L}}

\newcommand{\Pc}{{\mathcal P}}
\newcommand{\Qc}{{\mathcal Q}}

\newcommand{\Sc}{{\mathcal S}}
\newcommand{\Tc}{{\mathcal T}}

\newcommand{\Vc}{{\mathcal V}}

\newcommand{\dsf}{{\mathsf d}}

\newcommand{\psf}{{\mathsf p}}
\newcommand{\qsf}{{\mathsf q}}

\newcommand{\Ksf}{{\mathsf K}}

\renewcommand{\det}{{\hbox{det}}}

\theoremstyle{definition}
\newtheorem*{rem*}{Remark}

\theoremstyle{plain}
\newtheorem{thm}{Theorem}[section]

\newtheorem{lem}{Lemma}

\newtheorem{rem}{Remark}

\providecommand{\definitionname}{Definition}

\newcommand{\rk}{\mathsf{rk}}

\allowdisplaybreaks

\title{An Achievable Scheme for the K-user \\ Linear Computation Broadcast Channel}

\begin{document}

\author{
\IEEEauthorblockN{Yinbin Ma and Daniela Tuninetti}
\IEEEauthorblockA{University of Illinois Chicago, Chicago, IL 60607, USA \\ 
Email:\{yma52, danielat\}@uic.edu}
}

\maketitle

\IEEEpeerreviewmaketitle

\begin{abstract}
    This paper presents a new achievable scheme for the K-user Linear Computation Broadcast Channel (K-LCBC).
    A K-LCBC comprises data stored on a server and K users, each aiming to retrieve a desired linear function of the data by leveraging their prior locally available side information in the form of another linear function of the data.
    The proposed scheme is based on a %
    subspace decomposition derived from representable polymatroid spaces. 
    This decomposition enables the server to effectively design multicast messages that simultaneously benefit multiple users and allow users to eliminate interference using their available side information.
    This work extends existing results for the 3-LCBC by introducing a linear programming framework to optimize multicast opportunities across an arbitrary number of users. 
    The proposed approach can be used to derive achievable scheme for the K-user coded caching problem with linear coded placement and scalar linear function retrieval, which was our original motivation to investigate the K-LCBC.
    \end{abstract}

    \section{Introduction}
    \label{sec:intro}
    The Linear Computation Broadcast Channel (LCBC) was first introduced by Sun and Jafar in~\cite{sun2019capacity}. It is a canonical model motivated by distributed computing, where each %
    user requests some linear functions of the data stored at a server while having locally stored some other linear functions of the data. 
    The %
    {\it capacity} of the LCBC is defined as the minimum communication load from the server to the users that allows all users' demands to be satisfied.
    In~\cite{sun2019capacity} and~\cite{yao2024capacity}, the capacity of LCBC for 2 and 3 users, respectively, was characterized where the optimal scheme is derived from linear rank inequalities over a representable polymatroid space with no more than three variables~\cite{hammer2000inequalities}. 
    The capacity of the LCBC with four or more users is open. 
    The LCBC, as other network-coding-type problems, necessitate the characterization of the `entropy cone,' which represents all possible joint entropy relationships among variables in a network which involves the highly non-trivial problem of identifying its extreme rays and facets. The entropic approach, which was sufficient for the 2-LCBC~\cite{sun2019capacity}, already failed for the 3-LCBC even with Shannon and non-Shannon information inequalities included~\cite{yao2024capacity}. The tight converse bound in~\cite{yao2024capacity} was derived by also using `functional sub-modularity' to account for the specific functional forms of users' demand and side information. 
    As the general LCBC problem appears beyond reach,~\cite{yao2024generic} pursued a {\it generic and symmetric} version of the LCBC problem for which the capacity was characterized to within a factor of two.
    
    Despite those difficulties, we are motivated to understand the ultimate performance of the general K-user LCBC, as the LCBC model encompasses index coding as a special case; we note that index coding was instrumental in understanding the ultimate performance limits of coded caching with uncoded placement~\cite{wan2020index}. Similarly, the LCBC could potentially be leveraged to derive ultimate performance results for coded caching with linear coded placement, which is at present open.

    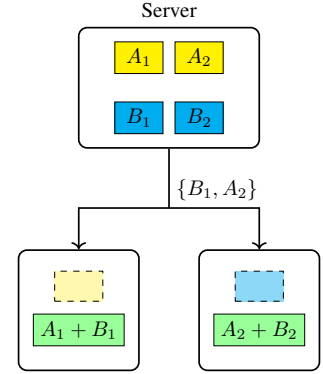
\begin{figure}
        \centering
        \scalebox{0.785}{
        \begin{tikzpicture}
            \node [draw, rounded corners, thick, minimum width=3cm, minimum height=2cm] (server) at (2.5,5.5) {};
            \node[left=1pt of server] {Server};
            \node [draw, minimum width=0.8cm, minimum height=0.5cm, fill=yellow] (A1) at (2, 6) {$A_1$};
            \node [draw, minimum width=0.8cm, minimum height=0.5cm, fill=cyan] (B1) at (2, 5) {$B_1$};
            \node [draw, minimum width=0.8cm, minimum height=0.5cm, fill=yellow] (A2) at (3, 6) {$A_2$};
            \node [draw, minimum width=0.8cm, minimum height=0.5cm, fill=cyan] (B2) at (3, 5) {$B_2$};
            \node [draw, minimum width=2cm, minimum height=2cm, thick, rounded corners] (U1) at (1,1.8) {};
            \node [draw, minimum width=2cm, minimum height=2cm, thick, rounded corners] (U2) at (4,1.8) {};
            \draw[->, thick] (2.5, 3.5) -| (U1);
            \draw[->, thick] (2.5, 3.5) -| (U2);
            \draw[thick] (2.5, 3.5) -- (server);
            \node [draw, dashed, minimum width=0.8cm, minimum height=0.5cm, fill=yellow!40] (D1) at (1,2.2) {};
            \node [draw, dashed, minimum width=0.8cm, minimum height=0.5cm, fill=cyan!40] (D2) at (4,2.2) {};
            \node [draw, minimum width=0.8cm, minimum height=0.5cm, below=0.2cm of D1, fill=green!40] {$A_1+B_1$};
            \node [draw, minimum width=0.8cm, minimum height=0.5cm, below=0.2cm of D2, fill=green!40] {$A_2+B_2$};
            \node [above right] at (2.5, 3.5) {$\{B_1, A_2\}$};
        \end{tikzpicture}
        }
        \caption{\small Optimal 2-LCBC scheme, %
        where user~1 wants $A_\star$ and knows $A_1 + B_1$, and  where user~2 wants $B_\star$ and knows $A_2 + B_2$. }
        \label{fig:LCBClayout}
    \end{figure}

    The LCBC is closely related to various problems in information theory, such as coded caching~\cite{maddah2014fundamental} and index coding~\cite{arbabjolfaei2018fundamentals}.
    Fig.~\ref{fig:LCBClayout} illustrates a 2-LCBC example.
    It can be seen as an index coding problem, where the user~1 (resp.~2) demands $A_1, A_2$ (resp. $B_1, B_2$) with side information $A_1 + B_1$ (resp. $A_2 + B_2$). In this example, the users cannot infer any piece of their desired information from what they locally have as side information. 
    The optimal scheme is to send $B_1$ and $A_2$, which serves two users simultaneously with load of one file. 
    It is also can be seen as a coded caching problem\footnote{Coded caching with linear coded placement and linear function retrieval is different from the LCBC problem. In the LCBC, the linear functions representing the users' demands and cached contents are fixed and known in advance. In contrast, in coded caching, the cache contents must be carefully designed ahead of time and remain fixed, while the system must guarantee good performance in a worst-case sense over all possible user demands.}, where $A_\star$ and $B_\star$ (where each file has been split into two equal-size subfiles) are the two files demanded by the users.

    \paragraph*{Contributions}
    Inspired by~\cite{yao2024capacity}, we introduce a general achievable scheme for the K-LCBC with a general number of users, by solving the linear programming problem in Theorem~\ref{thm:achievabilityLCBCK}, which is based on a general subspace decomposition derived from representable polymatroid spaces. The novelty of the scheme is to leverage linear dependency among subspaces related to information demanded and cached by users, making multicast and interference elimination possible for load saving.

    \paragraph*{Paper Organization}
    Section~\ref{sec:model} introduces the K-LCBC model. 
    Section~\ref{sec:generalsubspacedecomposition} explains the general subspace decomposition. 
    Section~\ref{sec:achievabilityLCBCK} introduces the proposed achievable scheme for general $K$. 
    Section~\ref{sec:examples} provides an examples of our proposed scheme for $K=4$ users. 
    Section~\ref{sec:conclusion} concludes this paper. 
    Further proofs for Lemmas and Theorems are provided in Appendix.

    \paragraph*{Notation Convention}
        Calligraphic symbols denote sets, bold lowercase symbols %
        vectors, bold uppercase symbols matrices, and sans-serif symbols %
        system parameters.
        $|\cdot|$ is the cardinality of a set or the length of a vector.
        For integers $a$ and $b$, $\binom{a}{b}$ is the binomial coefficient, or 0 if $a \geq b \geq 0$ does not hold.
        For integers $a \leq b$, we let $[a: b] := \{a, a+1, \ldots, b\}$ and $[b] := [1: b]$. %
        For sets $\Sc$ and $\Qc$, we let $\Sc \setminus \Qc := \{k: k \in \Sc, k \notin \Qc\}$.
        We denote the the collection of all subsets from set $\Sc$ with cardinality $t$ as $\Omega_{\Sc}^t$. 
        For example, $\Omega_{[3]}^2 = \{\{1,2\}, \{1,3\}, \{2,3\}\}$.

    \section{System Model}
    \label{sec:model}
    
    Our adopted notations follows~\cite{yao2024capacity}, however some %
    variables have been renamed to align with the notation conventions in coded caching.
    The K-LCBC is defined as follows for given $(\qsf, \Ksf, \dsf, (m_k, m'_k)_{k \in [\Ksf]}, (\Vm_k, \Vm'_k)_{k \in [\Ksf]})$.
    \begin{itemize}
    
        \item A server stores a data stream $\fv(t), \forall t \in \NN$, where 
        \begin{align*}
        \fv(t) = [f_1(t), f_2(t), \ldots, f_\dsf(t)] \in \FF_\qsf^{\dsf \times 1},
        \end{align*}
        where $f_i(t), \forall i \in  [\dsf]$ is a i.i.d. uniform random variable on $\FF_\qsf$.
        
        \item $\Ksf$ users are connected to the server via error-free shared link. 
        Each user $k \in [\Ksf]$ has side information 
        \begin{align*}
            \wv'_k(t) = \fv(t) \Vm'_k \in \FF_\qsf^{1 \times m'_k}, \text{ where } \Vm'_k \in \FF_\qsf^{\dsf \times m'_k}.
        \end{align*}
        
        \item Each user $k \in [\Ksf]$ demands %
        \begin{align*}
            \wv_k(t) = \fv(t) \Vm_k \in \FF_\qsf^{1 \times m_k}, \text{ where } \Vm_k \in \FF_\qsf^{\dsf \times m_k}.
        \end{align*}
    \end{itemize}
    WLOG, we assume that $\Vm'_k$ and $\Vm_k$ are linearly independent for all $k \in [\Ksf]$, otherwise, some desired information is already in the local cache, in other words $\rk([\Vm'_k, \Vm_k]) = \rk(\Vm'_k) + \rk(\Vm_k) = m'_k+m_k, \forall k\in[\Ksf]$~\cite{yao2024capacity}.
    An achievable $(L, N, \phi, (\varphi_k)_{k \in [\Ksf]})$ coding scheme consists of the following.
    \begin{itemize}
    
        \item The server aggregates $L$ instances of the data as
        \begin{align}
            \Fm = [\fv(1), \ldots, \fv(L)] \in \FF_\qsf^{\dsf \times L}.
        \end{align}
        For every user $k \in [\Ksf]$, the aggregated side information and the desired information are
        \begin{align}
            \Wm_k &= [\wv_k(1), \ldots, \wv_k(L)] \in \FF_\qsf^{L \times m_k}, \\
            \Wm'_k &= [\wv'_k(1), \ldots, \wv'_k(L)] \in \FF_\qsf^{L \times m'_k}.
        \end{align}
        
        \item The server uses the encoding function $\phi: \FF_\qsf^{\dsf \times L} \rightarrow \FF_\qsf^{N}$, and sends the message $X = \phi(\Fm)$ to the users.
        
        \item User $k \in [\Ksf]$, given decoding function $\varphi_k: \FF_\qsf^{L \times m'_k} \times \FF_\qsf^{N} \rightarrow \FF_\qsf^{L \times m_k}$, recovers $\Wm_k$ as $\Wm_k = \varphi_k(\Wm'_k, X)$. 
        
        \item The rate of the coding scheme is defined as $R = N/L$. 
    \end{itemize}
    Let the set of all achievable LCBC coding scheme be denoted by $\mathfrak{C}$. 
    The LCBC capacity is
    $%
        C = \inf_{(L, N, \phi, (\varphi_k)_{k \in [\Ksf]}) \in \mathfrak{C}} N/L
    $%
    {\footnote{This definition of capacity is for fixed $(\Vm_k, \Vm'_k)_{k \in [\Ksf]}$. 
    The generic LCBC capacity is defined as follows:
    let $\qsf = \psf^n$ where $\psf$ is a prime number and $n$ is a positive integer;
    fix $\Ksf, \dsf$ and $(m_k, m'_k)_{k \in [\Ksf]}$; 
    let $\Lc_n$ collect all instances of LCBC with parameters $({\psf^n}, \Ksf, \dsf, (m_k, m'_k)_{k \in [\Ksf]})$.
    \cite{yao2024generic} derives the capacity of the generic LCBC, i.e., the capacity of almost all LCBC instances in $\Lc_n$ when $n\to\infty$. 
    Due to space limit we omit details of generic capacity here.}}.

    Next we shall derive a scheme for the K-LCBC in Theorem~\ref{thm:achievabilityLCBCK} that recovers~\cite[Theorem~2]{sun2019capacity} and~\cite[Theorem~2]{yao2024capacity} as special cases for two and three users, respectively.

    \section{General Subspace Decomposition}
    \label{sec:generalsubspacedecomposition}
    
    Before we present our main results, we introduce important preliminary background. 
    We denote the rank of a matrix $\Mm$ as $\rk(\Mm)$. 
    If all columns of $\Mm$ are linear independent, we denote the subspace spanned by $\Mm$ as $\langle \Mm \rangle$ and we refer to $\Mm$ as its basis. 
    Given two matrices $\Mm_1$ and $\Mm_2$ from the same linear space, the conditional rank is defined as
    \begin{align}
        \rk(\Mm_1 | \Mm_2) := \rk([\Mm_1, \Mm_2]) - \rk(\Mm_2).
    \end{align}
    
    Let $\Tc$ be a ground set and  $(\Mm_{k})_{k \in \Tc}$ be a collection of matrices.
    For a subset $\Sc \subseteq \Tc$, we denote the union and intersection of matrices as follows
    \begin{align}
        \Mm_{\Sc} := \bigcap_{k \in \Sc} \langle\Mm_{k}\rangle, \quad
        \Mm_{(\Sc)} := \bigcup_{k \in \Sc} \langle\Mm_{k}\rangle,
        \label{eq:intersectionandunionsubspaces}
    \end{align}
    that is, the basis $\Mm_{\Sc}$ %
    is contained by all $\langle\Mm_k\rangle_{k \in \Sc}$, and 
    $\Mm_{(\Sc)}$ is the basis %
    of $\langle[\Mm_k: k \in \Sc]\rangle$. %
    For examples, $\Mm_{123} := \langle\Mm_1\rangle \cap  \langle\Mm_2\rangle \cap  \langle\Mm_3\rangle$
    \footnote{To keep notation light, we use $123$ as a subscript to denote the set $\{1,2,3\}$; we often adopt this notation when the context is clear.},
    that is, $\langle \Mm_{123} \rangle \subseteq \langle \Mm_{k} \rangle, \forall k \in [3]$;
    $\Mm_{(123)} := \langle\Mm_1\rangle \cup  \langle\Mm_2\rangle \cup  \langle\Mm_3\rangle$, that is,
    $\langle \Mm_{(123)} \rangle = \langle [\Mm_1, \Mm_2, \Mm_3] \rangle$.

    Given $\Sc \subseteq \Tc$ and $k \in \Tc \setminus \Sc$, we define %
    \begin{align}
        \Mm_{k(\Sc)} := \langle\Mm_k\rangle \cap \ \Big(\bigcup_{\ell \in \Sc} \langle\Mm_{\ell}\rangle \Big),
        \label{eq:compositionsubspace}
    \end{align}
    for example, $\Mm_{1(23)} := \langle\Mm_1\rangle \cap \big(\langle\Mm_2 \rangle \cup \langle \Mm_3 \rangle \big)$.
    
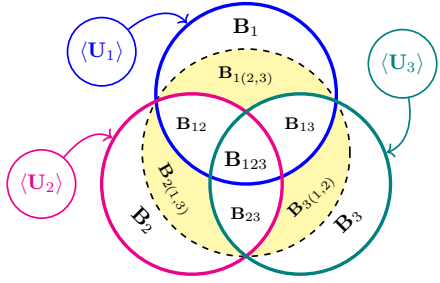
\begin{figure}
        \center
        \scalebox{0.8}{
        \begin{tikzpicture}
        \node [draw,
            circle,
            minimum size =3cm, color = olive, thick] (U1) at (0.9,1.5){};
        \node [draw,
            circle,
            minimum size =3cm, color = magenta, thick] (U2) at (0,0){};
        \node [draw,
            circle,
            minimum size =3cm, color = teal, thick] (U3) at (1.8,0){};
        \node [draw, dashed,
            circle,
            minimum size =3.45cm, fill = yellow!40, thick] (D) at (0.9,0.52){};
        \node [draw,
            circle,
            minimum size =0.5cm, color = blue, thick] (UU1) at (-1.5,2.2){$\langle {\Um_1} \rangle$};
        \node [draw,
            circle,
            minimum size =0.5cm, color = magenta, thick] (UU2) at (-2.5,0){$\langle {\Um_2} \rangle$};
        \node [draw,
            circle,
            minimum size =0.5cm, color = teal, thick] (UU3) at (3.5,2){$\langle {\Um_3} \rangle$};
            
        \begin{scope}
            \clip (0,0) circle(1.5cm);
            \clip (0.9,1.5) circle(1.5cm);
            \fill[white](0,0) circle(1.5cm);
        \end{scope}
        
        \begin{scope}
            \clip (0,0) circle(1.5cm);
            \clip (1.8,0) circle(1.5cm);
            \fill[white](0,0) circle(1.5cm);
        \end{scope}
        
        \begin{scope}
           \clip (0.9,1.5) circle(1.5cm);
            \clip (1.8,0) circle(1.5cm);
            \fill[white](0.9,1.5) circle(1.5cm);
        \end{scope}
        
        \node at (0.9,0.4) {$\Bm_{123}$};
        \node at (0,1) {\footnotesize $\Bm_{12}$};
        \node at (1.8,1) {\footnotesize $\Bm_{13}$};
        \node at (0.9,-0.5) {\footnotesize $\Bm_{23}$};
        \node at (0.9,1.8) {\footnotesize {$\Bm_{1(2,3)}$}};
        \node[rotate=-60] at (-0.3,-0.1) {\footnotesize $\Bm_{2(1,3)}$};
        \node[rotate=45] at (2,-0.2) {\footnotesize $\Bm_{3(1,2)}$};
        \node at (0.9,2.6) {$\Bm_{1}$};
        \node[rotate=-45] at (-0.8,-0.6) {$\Bm_{2}$};
        \node[rotate=45] at (2.6,-0.6) {$\Bm_{3}$};
        
        \node [draw,
            circle,
            minimum size =3cm, color = blue, ultra thick] (U1) at (0.9,1.5){};
        \node [draw,
            circle,
            minimum size =3cm, color = magenta, ultra thick] (U2) at (0,0){};
        \node [draw,
            circle,
            minimum size =3cm, color = teal, ultra thick] (U3) at (1.8,0){};
        \draw [->, blue, thick] (UU1) to [out=40, in=130](U1);
        \draw [->, teal,  thick] (UU3) to [out=-80, in=20](U3);
        \draw [->, magenta, thick] (UU2) to [out=50, in=150](U2);
        
        \end{tikzpicture}
        }
        \caption{\small The decomposition of $\langle \Um_1\rangle, \langle \Um_2\rangle, \langle \Um_3\rangle$ into subspaces, labeled by corresponding bases, from~\cite{yao2024capacity}.}
        \label{fig:venn1}
    \end{figure}

    Next we introduce a subspace decomposition for any $\Ksf$ matrices inspired by~\cite{yao2024capacity}.
    Given matrices  $\{ \Um_{k} : {k \in [\Ksf]} \}$
    that are full column rank (so that the matrix columns form a basis for the corresponding subspace, allowing us to avoid using angular brackets to denote the subspace spanned by the matrix columns), 
    there exists a collection of subspaces $\Vc := \Vc_1 \cup \Vc_2 \cup \Vc_3$ where
    \begin{align}
        \Vc_1 &= \{\Um_{\Sc}: \Sc \subseteq [\Ksf], |\Sc| \geq 2\}, 
        \label{eq:firsttypespace} \\
        \Vc_2 &= \{\Um_{k(\Tc \setminus {k})}: \Tc \subseteq [\Ksf], |\Tc| \geq 3, k \in \Tc\}, 
        \label{eq:secondtypespace}\\
        \Vc_3 &= \{\Um_k, k \in [\Ksf]\}. 
        \label{eq:lasttypespace}
    \end{align}
    For example, when $\Ksf=3$, we have
    \begin{align*}
        \Vc_1 &= \{ \Um_{12}, \Um_{13}, \Um_{23}, \Um_{123} \}, \\
        \Vc_2 &= \{ \Um_{1(23)}, \Um_{2(13)}, \Um_{3(12)} \}, \\
        \Vc_3 &= \{ \Um_{1},  \Um_{2},  \Um_{3} \}.
    \end{align*}
    For a subspace $V \in \Vc$, we define a function $\mathsf{LS}(V)$ as
    \begin{itemize}
        \item if $V = \Um_{[\Ksf]}$, then $\mathsf{LS}(V) = \emptyset$;
        \item if $V \in \Vc_1 \setminus \{ \Um_{[\Ksf]} \}$, i.e., $V = \Um_{\Sc} : 2\leq |\Sc| \leq \Ksf-1$, 
        \begin{align} 
            \mathsf{LS}(V) = \{\Um_{\Sc \cup \{\ell\}}: \ell \in [\Ksf] \setminus \Sc\};
        \end{align}
        \item if $V \in \Vc_2$, that is $V = \Um_{k(\Sc)}$, then 
        \begin{align}
            \mathsf{LS}(V) = \{\Um_{k(\Sc \setminus {\ell})}: \ell \in \Sc \};
        \end{align}
        \item if $V \in \Vc_3$, that is $V = \Um_k$, then 
        \begin{align}
            \mathsf{LS}(V) = \{\Um_{k([\Ksf]\setminus\{k\})}\}.
        \end{align}
    \end{itemize}
    For example, when $\Ksf=3$,
    $
    \mathsf{LS}(\Um_{12}) = \{ \Um_{123} \}, \mathsf{LS}(\Um_{1(23)}) = \{ \Um_{12}, \Um_{13} \}, \mathsf{LS}(\Um_{1}) = \{ \Um_{1(23)} \}.
    $
    We define the basis that spans the subspace $V$ only, but not included in $\mathsf{LS}(V)$, as 
    \begin{align}
        \Bm(V) := V \setminus \bigcup_{Q \in \mathsf{LS}(V)} Q. \label{eq:basisdefinition}
    \end{align}
    For example, when $\Ksf=3$, Fig.~\ref{fig:venn1} lists all possible subspaces and bases from~\cite[Lemma~2]{yao2024capacity}, where, in order to keep notation simple, we let $\Bm_{\star} = \Bm(\Um_{\star})$.
    Finally, we recursively define $\mathsf{C}(V)$ as the set of bases covered by subspace $V$ as
    \begin{align}
        \mathsf{C}(V) = \Bm(V) \cup \Big(\bigcup_{U \in \mathsf{LS}(V)} \mathsf{C}(U) \Big). 
        \label{eq:definitionofCfunciton}
    \end{align}
    
    Note that we do not consider here subspaces more `complex' than those in~\eqref{eq:secondtypespace}, such as %
    $
     \Um_{(12)(34)}=
     \big(\Um_1 \cup \Um_2\big)
    \cap
     \big(\Um_3 \cup \Um_4\big)
    $,
    which may arise when considering $\Ksf\geq 4$ users.
    This omission %
    simplifies the generalization to an arbitrary number of $\Ksf$ users, which is the focus of our current work. Ongoing research is aimed at understanding how this choice impacts the performance of the achievable scheme.

    \begin{lem}\cite[Lemma~2]{yao2024capacity} \label{lem:decompositionK3}\rm 
        $\{\Bm_{1(23)}, \Bm_{2(13)}, \Bm_{3(12)}\}$ have identical dimension, and they are pairwise linearly independent.
    \end{lem}
    From Lemma~\ref{lem:decompositionK3},~\cite{yao2024capacity} shows the subspace decomposition for any $\Ksf=3$ matrices as in Fig.~\ref{fig:venn1}, where the yellow bases are mutually independent with identical size -- akin to an MDS code of rate $2/3$.
    Now we %
    generalize Lemma~\ref{lem:decompositionK3} to arbitrary $\Ksf$.
    \begin{lem} \label{lem:decompositionK} \rm
        Given any $\Tc \subseteq [\Ksf]$ and $ t :=|\Tc| \geq 3$, 
        $\{\Bm_{k(\Tc \setminus \{k\})}: k \in \Tc\}$ have identical dimension, and any $t-1$ of them are linearly independent. 
    \end{lem}
    We provide the proof of Lemma~\ref{lem:decompositionK} in Appendix~\ref{sec:proofdecompositionK}, and we include further remarks in Appendix~\ref{sec:remarklemmadecompositionK}.

    \section{Achievable scheme for the K-LCBC}
    \label{sec:achievabilityLCBCK}
    An achievable rate for the K-LCBC is derived as the solution of a Linear Program (LP). 
    To do that, first we need to specify our optimizing variables, whose essence is coding gain from linear coding.
    Given any $\Ksf$ matrices, we assign a coefficient to each basis. Specifically, for every $\Um_{\Sc} \in \Vc_1$ defined in~\eqref{eq:firsttypespace}, we assign $\lambda_\Sc$ to its base $\Bm_\Sc$; for every $\Um_{k(\Sc)}$ defined in~\eqref{eq:secondtypespace}, we assign $\lambda_{(\{k\} \cup \Sc)}$ to its base $\Bm_{k(\Sc)}$; for every $\Um_k$ defined in~\eqref{eq:lasttypespace}, we assign $\lambda_k$ to its base $\Bm_k$.
    Fig.~\ref{fig:venn2} shows an example of coefficient assignments for $\Ksf = 3$.
    We explain later on the reason why the yellow subspaces share the same coefficient $\lambda_{(123)}$.
    
    For every subspace $V \in \Vc$, we define the function $\mathsf{C}_\lambda(V)$ that returns the set of coefficients covered by $\mathsf{C}(V)$ defined in~\eqref{eq:definitionofCfunciton}. %
    For example, for $\Ksf=3$, we have
    \begin{align*}
    \mathsf{C}_\lambda(\Um_{12}) &= \{\lambda_{12}, \lambda_{123}\}, \\
    \mathsf{C}_\lambda(\Um_{1(2,3)}) &= \{\lambda_{(123)}, \lambda_{12}, \lambda_{13}, \lambda_{123}\}, \\ 
    \mathsf{C}_\lambda(\Um_{1}) &= \{\lambda_{1},\lambda_{(123)}, \lambda_{12}, \lambda_{13}, \lambda_{123}\}.
    \end{align*}
    
    \begin{figure}
        \center
        \scalebox{0.8}{
        \begin{tikzpicture}
        \node [draw,
            circle,
            minimum size =3cm, color = olive, thick] (U1) at (0.9,1.5){};
        \node [draw,
            circle,
            minimum size =3cm, color = magenta, thick] (U2) at (0,0){};
        \node [draw,
            circle,
            minimum size =3cm, color = teal, thick] (U3) at (1.8,0){};
        \node [draw, dashed,
            circle,
            minimum size =3.45cm, fill = yellow!40, thick] (D) at (0.9,0.52){};
        \node [draw,
            circle,
            minimum size =0.5cm, color = blue, thick] (UU1) at (-1.5,2.2){$\langle {\Um_1} \rangle$};
        \node [draw,
            circle,
            minimum size =0.5cm, color = magenta, thick] (UU2) at (-2.5,0){$\langle {\Um_2} \rangle$};
        \node [draw,
            circle,
            minimum size =0.5cm, color = teal, thick] (UU3) at (3.5,2){$\langle {\Um_3} \rangle$};
            
        \begin{scope}
            \clip (0,0) circle(1.5cm);
            \clip (0.9,1.5) circle(1.5cm);
            \fill[white](0,0) circle(1.5cm);
        \end{scope}
        
        \begin{scope}
            \clip (0,0) circle(1.5cm);
            \clip (1.8,0) circle(1.5cm);
            \fill[white](0,0) circle(1.5cm);
        \end{scope}
        
        \begin{scope}
           \clip (0.9,1.5) circle(1.5cm);
            \clip (1.8,0) circle(1.5cm);
            \fill[white](0.9,1.5) circle(1.5cm);
        \end{scope}
    
        \node at (0.9,0.4) {\footnotesize $\lambda_{123}$};
        \node at (0,1) {\footnotesize $\lambda_{12}$};
        \node at (1.8,1) {\footnotesize $\lambda_{13}$};
        \node at (0.9,-0.5) {\footnotesize $\lambda_{23}$};
        \node at (0.9,1.8) {\footnotesize {$\lambda_{(1,2,3)}$}};
        \node[rotate=-60] at (-0.3,-0.1) {\footnotesize $\lambda_{(1,2,3)}$};
        \node[rotate=45] at (2,-0.2) {\footnotesize $\lambda_{(1,2,3)}$};
        \node at (0.9,2.6) {$\lambda_{1}$};
        \node[rotate=-45] at (-0.8,-0.6) {$\lambda_{2}$};
        \node[rotate=45] at (2.6,-0.6) {$\lambda_{3}$};

        \node [draw,
            circle,
            minimum size =3cm, color = blue, ultra thick] (U1) at (0.9,1.5){};
        \node [draw,
            circle,
            minimum size =3cm, color = magenta, ultra thick] (U2) at (0,0){};
        \node [draw,
            circle,
            minimum size =3cm, color = teal, ultra thick] (U3) at (1.8,0){};
        \draw [->, blue, thick] (UU1) to [out=40, in=130](U1);
        \draw [->, teal,  thick] (UU3) to [out=-80, in=20](U3);
        \draw [->, magenta, thick] (UU2) to [out=50, in=150](U2);
        
        \end{tikzpicture}
        }
        \caption{\small Coefficients associated to each basis for $\Ksf=3$ users. 
        }
        \label{fig:venn2}
    \end{figure}

    \begin{thm} \rm \label{thm:achievabilityLCBCK}
        For the K-LCBC with side information $(\Vm'_k)_{k \in [\Ksf]}$ and desired information $(\Vm_k)_{k \in [\Ksf]}$, let $\Um_k = [\Vm'_k; \Vm_k]$.
        The following rate is achievable %
        \begin{align} 
        \min_{\lambda_\star \geq 0} \
        \sum_{\Sc \subseteq [\Ksf]} \lambda_\Sc + \sum_{t=3}^{\Ksf} \sum_{\Sc \in \Omega_{[\Ksf]}^{t}} (t-1) \lambda_{(\Sc)},
           \label{eq:objectivefunctionloadsum}
        \end{align}
        where $\lambda_{\star}$ are coefficients associated to bases for $\Vc_1$, $\Vc_2$ and $\Vc_3$.
        The constraints for $\lambda_{\star}$ are generated in three stages separately, namely%
        \begin{subequations}
        \begin{align}
            &\text{"Sum of $\mathsf{C}_\lambda(\Um_{\Sc})$"} \leq \rk(\Um_{\Sc} \mid \Vm'_k), \label{eq:firsttypeconstraintmain}\\ 
            &\text{"Sum of $\bigcup_{V \in \Lc_\Sc} \mathsf{C}_\lambda(V)$"} \leq \rk(\Lc_\Sc \mid \Vm'_k), \label{eq:firsttypeconstraintsubset}
        \end{align}
        \label{eq:firsttypeconstraint}
        \end{subequations}
        where $\Lc_\Sc \subseteq \mathsf{LC}(\Um_\Sc)$, for every $\Um_\Sc \in \Vc_1$ and $k \in \Sc$; and %
        \begin{subequations}
        \begin{align}
            &\text{"Sum of $\mathsf{C}_\lambda(\Um_{k(\Sc)}$)"} \leq \rk(\Um_{k(\Sc)} \mid \Vm'_k), \label{eq:secondtypeconstraintsmain}\\ 
            &\text{"Sum of $\bigcup_{V \in \Lc_{k(\Sc)}} \mathsf{C}_\lambda(V)$"} \leq \rk(\Lc_{k(\Sc)} \mid \Vm'_k), \label{eq:secondtypeconstraintssubset}
        \end{align}
        \label{eq:secondtypeconstraints}
        \end{subequations}
        where $\Lc_{k(\Sc)} \subseteq \mathsf{LC}(\Um_{k(\Sc)})$, for every $\Um_{k,\Sc} \in \Vc_2$; and %
        \begin{align}
            \text{"Sum of $\mathsf{C}_\lambda(\Um_{k}$)"} = \rk(\Um_{k} \mid \Vm'_k) = \rk(\Vm_{k}), \label{eq:lasttypeconstraint}
        \end{align}
        for all $\Um_k \in \Vc_3$.
        $\hfill\square$
    \end{thm}

    \begin{rem} \label{rem:1} %
        From the equality in~\eqref{eq:lasttypeconstraint}, we can write~\eqref{eq:objectivefunctionloadsum} as
        \begin{align}
    \sum_{k=1}^{\Ksf} \rk(\Um_k | \Vm'_k) 
    - \sum_{t=2}^{\Ksf} \sum_{\Sc \in \Omega_{[\Ksf]}^{t}} (t-1) \lambda_\Sc 
    - \!\!\!\!
    \sum_{\Sc \subseteq [\Ksf], |\Sc| \geq 3} 
    \!\!\!
    \lambda_{(\Sc)}. \label{eq:objectivefunctioncodinggain}
        \end{align}
        In~\eqref{eq:objectivefunctioncodinggain}, the first term is the load of uncoded transmissions when users are served individually. 
        Each $\lambda_{\star}$ can be viewed as the coding gain from a multicast message transmission. 
        The goal is to find multicast messages that satisfy multiple users simultaneously. 
        The constraints in~\eqref{eq:firsttypeconstraint} implies that
        there exists a message with rank $\lambda_{\Sc}$, which serves $t=|\Sc|$ of users, thus the load is reduced by $(t-1)\lambda_\Sc$ as shows in~\eqref{eq:objectivefunctioncodinggain}.
        The constraints in~\eqref{eq:secondtypeconstraints} implies that, for every $t \geq 3$ and $\Sc \in \Omega_{[\Ksf]}^{t}$ (i.e., any $t$ users indexed by $\Sc$),
        there exist $t$ messages, of which $t-1$ are linearly independent, to serve the $t$ of users. Each message has the same rank $\lambda_{(\Sc)}$, thus the load is reduced by $\lambda_{(\Sc)}$ as shown in~\eqref{eq:objectivefunctioncodinggain}.
        The constraint in~\eqref{eq:lasttypeconstraint} implies that the server sends unicast messages to `cover' the whole demand information space $\Vm_k$, for every $k\in[\Ksf]$, if they were not transmitted in previous multicast messages; no load saving are possible with unicast messages. 
        $\hfill\square$
    \end{rem}
    
    The proof of Theorem~\ref{thm:achievabilityLCBCK} is provided in Appendix~\ref{sec:proofachievabilityLCBCK}, where we show the existence of the multicast messages discussed in Remark~\ref{rem:1}, as well as show decoding correctness. 
    
    Section~\ref{sec:examples} provides a detailed example of Theorem~\ref{thm:achievabilityLCBCK} for $\Ksf=4$ users.
    When $\Ksf=2$ or $3$, Theorem~\ref{thm:achievabilityLCBCK} reduces to~\cite[Theorem~2]{sun2019capacity} or~\cite[Theorem~2]{yao2024capacity} respectively, which both achieves capacity.

    \section{Example for Theorem~\ref{thm:achievabilityLCBCK}}
    \label{sec:examples}
    
    In this section, we explain how to formulate the LP in Theorem~\ref{thm:achievabilityLCBCK} when $\Ksf=4$. 
    We only consider the perspective of user~1; similar reasoning applies to the remaining users.
    
    \subsection{Linear Programming when $\Ksf = 4$}
    \label{sec:exampleK=3}
    
    \paragraph*{Stage~1} 
    Let $t=4$.
    We start from %
    $\Um'_{1234}$ -- a submatrix of $\Um_{1234}$ that benefits all users\footnote{Based on those bases / submatrices, we shall construct multicast messages to benefit the users indicated in the subscript.}.
    It must satisfy
    the following constraint from~\eqref{eq:firsttypeconstraintmain}
    \begin{align}
        \lambda_{1234} \leq \rk(\Um_{1234} | \Vm'_1). \label{eq:multicast1234}
    \end{align}
    The RHS of~\eqref{eq:multicast1234} is an upper bound on the rank of $\Um'_{1234}$. 
    As $\mathsf{LS}(\Um_{1234}) = \emptyset$, the constraint from~\eqref{eq:firsttypeconstraintsubset} does not exist.

    Let $t=3$.
    We consider %
    $\{\Um'_{123}, \Um'_{124}, \Um'_{134}\}$ -- submatrices of
    $\{\Um_{123}, \Um_{124}, \Um_{134}\}$ that benefit three users at once.
    They must satisfy
    the following constraints from~\eqref{eq:firsttypeconstraintmain} 
    \begin{align}
        \lambda_{1234} + \lambda_{123} &\leq \rk(\Um_{123} | \Vm'_1), \label{eq:multicast123} \\
        \lambda_{1234} + \lambda_{124} &\leq \rk(\Um_{124} | \Vm'_1), \label{eq:multicast124} \\
        \lambda_{1234} + \lambda_{134} &\leq \rk(\Um_{134} | \Vm'_1), \label{eq:multicast134}
    \end{align}
    The RHSs of~\eqref{eq:multicast123}-\eqref{eq:multicast134} are upper bound on the rank of $\Um'_{123}$, $\Um'_{124}$, $\Um'_{134}$, respectively.
    Note that $\mathsf{LS}(\Um_{123}) = \mathsf{LS}(\Um_{124}) = \mathsf{LS}(\Um_{134}) = \mathsf{LS}(\Um_{1234})$, thus their constraints from~\eqref{eq:firsttypeconstraintsubset} are equivalent to~\eqref{eq:multicast1234}.
    
    Let $t=2$.
    We consider %
    $\Um'_{12}, \Um'_{13}, \Um'_{14}$ -- submatrices of $\Um_{12}, \Um_{13}, \Um_{14}$ that benefit pairs of users.
    They must satisfy
    the constraints from~\eqref{eq:firsttypeconstraintmain} 
    \begin{align}
        \lambda_{1234} + \lambda_{123} + \lambda_{124} + \lambda_{12} &\leq \rk(\Um_{12} | \Vm'_1), \label{eq:multicast12} \\
        \lambda_{1234} + \lambda_{123} + \lambda_{134} + \lambda_{13} &\leq \rk(\Um_{13} | \Vm'_1), \label{eq:multicast13} \\
        \lambda_{1234} + \lambda_{124} + \lambda_{134} + \lambda_{14} &\leq \rk(\Um_{14} | \Vm'_1). \label{eq:multicast14}
    \end{align}
    The RHSs of~\eqref{eq:multicast12}-\eqref{eq:multicast14} are upper bound on the rank of $\Um'_{12}, \Um'_{13}, \Um'_{14}$, respectively.
    We focus on $\Um'_{12}$, note that $\mathsf{LS}(\Um_{12}) = \{\Um_{123}, \Um_{124}\}$, the constraint~\eqref{eq:firsttypeconstraintsubset} is
    \begin{align}
        \lambda_{1234} + \lambda_{123} + \lambda_{124} &\leq \rk(\Um_{123}, \Um_{124} | \Vm'_1). \label{eq:multicast123,124}
    \end{align}
    The RHS of~\eqref{eq:multicast123,124} is an upper bounded on the rank $[\Um'_{123},\Um'_{124}]$.
    We proceed similarly for $\Um_{13}$ and $\Um_{14}$.

    \paragraph*{Stage~2} 
    Let $t=3$.
    We consider $\Um'_{1(23)}$, a submatrix of $\Um_{1(23)}$, whose rank is constrained by~\eqref{eq:secondtypeconstraintsmain} and~\eqref{eq:secondtypeconstraintssubset}; then
    \begin{align}
        \lambda_{1234} + \ldots + \lambda_{12} + \lambda_{13} + \lambda_{(123)} &\leq \rk(\Um_{1(23)} | \Vm'_1), \label{eq:multicast1,23}\\
        \lambda_{1234} + \ldots + \lambda_{12} + \lambda_{13} &\leq \rk(\Um_{12}, \Um_{13} | \Vm'_1). \label{eq:multicast1213} 
    \end{align}
    Similarly for $\Um_{1(24)}$ and $\Um_{1(34)}$.
    \cite{yao2024capacity} states that there exist $\Um'_{1(23)}, \Um'_{2(13)}, \Um'_{3(12)}$ with rank $\lambda_{(123)}$, which are submatries of $\Um_{1(23)}, \Um_{2(13)}, \Um_{3(12)}$, respectively, as shown in~\eqref{eq:multicast1,23}. 
    The RHS of~\eqref{eq:multicast1213} is an upper bound on the rank of $[\Um'_{12}, \Um'_{13}]$.
    
    Let $t=4$.
    We consider $\Um'_{(1234)}$, a submatrix of $\Um_{1(234)} \in \Vc_2$ whose rank is constrained by~\eqref{eq:secondtypeconstraintsmain}; then
    \begin{multline*}
        \lambda_{1234} + \ldots + \lambda_{(123)} + \lambda_{(124)} \\ + \lambda_{(134)} + \lambda_{(1234)} \leq \rk(\Um_{1(234)} | \Vm'_1). %
    \end{multline*}
    For~\eqref{eq:secondtypeconstraintssubset}, as $\mathsf{LS}(\Um_{1(234)}) = \{\Um_{1(23)}, \Um_{1(24)}, \Um_{1(34)}\}$, we get
    \begin{align*}
        &\lambda_{1234} + \ldots + \lambda_{(123)} + \lambda_{(124)} \leq \rk(\Um_{1(23)}, \Um_{1(24)} | \Vm'_1), %
        \\ &\lambda_{1234} + \ldots + \lambda_{(123)} + \lambda_{(134)} \leq \rk(\Um_{1(23)}, \Um_{1(34)} | \Vm'_1), %
        \\ &\lambda_{1234} + \ldots + \lambda_{(124)} + \lambda_{(134)} \leq \rk(\Um_{1(24)}, \Um_{1(34)} | \Vm'_1), %
        \\ &\lambda_{1234} + \ldots + \lambda_{(123)} + \lambda_{(124)} + \lambda_{(134)}
        \notag \\ &\quad \leq \rk(\Um_{1(23)}, \Um_{1(24)}, \Um_{1(34)} | \Vm'_1). %
    \end{align*}

    \paragraph*{Stage~3} 
    We consider $\Um_{1} \in \Vc_3$, from constraint~\eqref{eq:lasttypeconstraint}, we attain an unicast message which serves user~1 only, with
    \begin{align}
        \lambda_{1} + \lambda_{1234} + \ldots + \lambda_{(1234)} = \rk(\Um_1 | \Vm'_1).
    \end{align}

    \paragraph*{All together}
    In summary, the achievable rate is the sum of the ranks of all messages sent by server, that is,
    \begin{align}
        \sum_{\Sc \subseteq [4]} \lambda_{\Sc} + \sum_{t=3}^{4} \sum_{\Sc \in \Omega_{[4]}^t} (t-1) \lambda_{(\Sc)} \label{eq:capacityfor4users}
    \end{align}
    which is~\eqref{eq:objectivefunctionloadsum} when $\Ksf=4$ users.

    \subsection{ Construction of Multicast Messages} %

    \paragraph*{Stage 1} For every $V = \Um_\Sc \in \Vc_1$, we let $\Mm_{\Sc} \in \FF_\qsf^{\rk(\Um_\Sc) \times \lambda_{\Sc}}$, so
        $\Um'_{\Sc} = \Um_{\Sc} \Mm_{\Sc}$
    is a submatrix used to construct multicast message to benefit the users in~$\Sc$.
    For example, we have the following multicast messages:
    $\Um'_{1234} = \Um_{1234} \Mm_{1234}$,
    $\Um'_{123}  = \Um_{123} \Mm_{123}$, and 
    $\Um'_{12}   = \Um_{12} \Mm_{12}.$
    We omit others for the sake of space.
    
    \paragraph*{Stage 2} %
    Let $\Tc \subseteq [3]$ and $t := |\Tc| \geq 3$, we can always have %
    \begin{align}
        \sum_{k \in \Tc} \Um'_{k(\Tc \setminus \{k\})} = 0, \label{eq:sumofmulticastmessagesis0}
    \end{align} 
    where $\Um'_{k(\Tc \setminus \{k\})}$ is a submatrix of $\Um_{k(\Tc \setminus \{k\})}$.
    In~\cite[Section~8]{yao2024capacity}, this was proves for $t=3$.
    For $t=4$, we know $\Um_{1(234)}$ can be expressed as a linear span of 
    \begin{align*}
    \big[\Um_{12}, \Um_{13}, \Um_{14}, \Bm_{1(23)}, \Bm_{1(24)}, \Bm_{1(34)}, \Bm_{1(234)}\big].
    \end{align*}
    Similarly, $\Um_{2(134)}, \Um_{3(124)}, \Um_{4(123)}$ can be expressed, respectively, as a linear span of 
    \begin{align*}
        \big[\Um_{12}, \Um_{23}, \Um_{24}, \Bm_{2(13)}, \Bm_{2(14)}, \Bm_{2(34)}, \Bm_{2(134)}\big], \\
        \big[\Um_{13}, \Um_{23}, \Um_{34}, \Bm_{3(12)}, \Bm_{3(14)}, \Bm_{3(24)}, \Bm_{3(124)}\big], \\
        \big[\Um_{14}, \Um_{24}, \Um_{34}, \Bm_{4(12)}, \Bm_{4(13)}, \Bm_{4(23)}, \Bm_{4(123)}\big].
    \end{align*}
    Let %
    \begin{align*}
        &\Nm_{ij} \in \FF_\qsf^{\rk(\Um_{ij}) \times \lambda_{(1234)}} &, \ &\forall \{i,j\} \subseteq [4], \\
        &\Nm_{\Sc} \in \FF_\qsf^{\rk(\Bm_{k(\Sc \setminus \{k\})}) \times \lambda_{(1234)}} &, \ & \forall \Sc \subseteq [4], |\Sc| \geq 3, k \in \Sc.
    \end{align*}
    The multicast messages $\Um'_{1(234)}, \Um'_{2(134)}, \Um'_{3(124)}, \Um'_{4(123)}$ can be found by %
    \begin{align*}
        \Um'_{1(234)} &=
        \ \Um_{12} \Nm_{12} + \Um_{13} \Nm_{13} + \Um_{14} \Nm_{14} + \Bm_{1(23)} \Nm_{123} 
        \notag \\ &+ \Bm_{1(24)} \Nm_{124} + \Bm_{1(34)} \Nm_{134}+ \Bm_{1(234)} \Nm_{1234}, \\
        \Um'_{2(134)} &=
        -\Um_{12} \Nm_{12} + \Um_{23} \Nm_{23} + \Um_{24} \Nm_{24} + \Bm_{2(13)} \Nm_{123} 
        \notag \\ &+ \Bm_{2(14)} \Nm_{124} + \Bm_{2(34)} \Nm_{234}+ \Bm_{2(134)} \Nm_{1234}, \\
        \Um'_{3(124)} &= 
        -\Um_{13} \Nm_{13} - \Um_{23} \Nm_{23} + \Um_{34} \Nm_{34} - \Bm_{3(12)} \Nm_{123} 
        \notag \\ &+ \Bm_{3(14)} \Nm_{134} + \Bm_{3(24)} \Nm_{234}+ \Bm_{3(124)} \Nm_{1234}, \\
        \Um'_{4(123)} &= \
        \Um_{14} \Nm_{14} + \Um_{24} \Nm_{24} + \Um_{34} \Nm_{34} + \Bm_{4(12)} \Nm_{124} 
        \notag \\ &+ \Bm_{4(13)} \Nm_{134} + \Bm_{4(23)} \Nm_{234}+ \Bm_{4(123)} \Nm_{1234} \notag \\
        &\stackrel{(a)}{=} \Um'_{1(234)} + \Um'_{2(134)} + \Um'_{3(124)},
    \end{align*}
    where (a) follows from Lemma~\ref{lem:decompositionK}.
    These signs of each item are chosen with intention so~\eqref{eq:sumofmulticastmessagesis0} holds.
    The existence of matrices $\Nm$'s can be argued by drawing their coefficients independently uniformly at random over an extended finite field. By doing so, and by considering a sufficiently large field extension, one can argue that the probability of the system of matrix equations lacking a solution vanishes\footnote{We note that techniques mentioned in~\cite[Section 6]{yao2024capacity}, including vector coding and field extension, are necessary in generel to achieve rational rates / values for the lambda-parameters, which by the above description would appear to only take integer values.}. %
    
    \paragraph*{Stage 3}
    For every $k \in [\Ksf]$, let $\Mm_k \in \FF_\qsf^{\rk(\Um_k) \times \lambda_k}$; the server sends the unicast message $\Um'_k = \Um_k \Mm_k$, so that user~$k$ can decode the remaining information of $\Vm_k$ which was not transmitted in previous multicast messages.

    \paragraph*{All together}
    Thus, the multicast messages $X$ is
    \begin{multline*}
        X = \{\Fm \Um'_{\Sc}: \Sc \subseteq [4]\} \\ 
        \cup \Big(\bigcup_{t \geq 3} \text{"any $t-1$ of } 
        \{\Fm \Um'_{k(\Sc \setminus \{k\})}: \Sc \in \Omega_{[4]}^{t}\}\text{"}\Big).
    \end{multline*} 
    The achievable rate is
    as shown in~\eqref{eq:capacityfor4users}.

    \subsection{Toy Example}
    Examples in~\cite[Section 6]{yao2024capacity} illustrate the construction of multicast messages for $\Ksf=3$ users. 
    Here we give an index coding example for $\Ksf=4$ users, whose optimal rate is known to be $3$~\cite[Section~5.1]{arbabjolfaei2018fundamentals}. We use this known example here to showcase an example of 4-birds 3-stones multicast messages.

    Suppose that $\fv = [A,B,C,D]$;
    user~1 wants $B$, knows $A$; 
    user~2 wants $C$, knows $B$; 
    user~3 wants $D$, knows $C$; and
    user~4 wants $A$, knows $D$.
    The subspace decomposition yields
    \begin{align*}
        \Bm_{12} = B, \ \ \ \Bm_{23} = C, \ \ \ \Bm_{34} = D, \ \ \ \Bm_{14} = A,
    \end{align*}
    and other bases are empty. %
    
    \paragraph*{Stage~1}
    \eqref{eq:multicast1234}-\eqref{eq:multicast134} are trivial as they are all zeros.
    \eqref{eq:multicast12}-\eqref{eq:multicast14} become
    $
        \lambda_{12} \leq 1, \lambda_{13} \leq 0, \lambda_{14} \leq 0.
    $
    Furthermore, %
    \begin{align*}
        \lambda_{12} &\leq \rk(\Um_{12} | \Vm'_2) = \rk([B, B]) - \rk([B]) = 0, \\
        \lambda_{23} &\leq \rk(\Um_{23} | \Vm'_3) = \rk([C, C]) - \rk([C]) = 0, \\
        \lambda_{34} &\leq \rk(\Um_{34} | \Vm'_4) = \rk([D, D]) - \rk([D]) = 0.
    \end{align*}
    Thus, for $\Sc \subseteq [4]$ and $|\Sc| \geq 2$, $\lambda_\Sc = 0$, that is, no multicast messages have been found.
    
    \paragraph*{Stage~2}
    \eqref{eq:multicast1,23} yields $\lambda_{(123)} \leq 1$, however
    \begin{align*}
        \lambda_{(123)} &\leq \rk(\Um_{3(12)} | \Vm'_3) = \rk([C,C]) - \rk([C]) = 0,
    \end{align*}
    which implies $\lambda_{(123)} = 0$. Similarly 
    \begin{align*}
        \lambda_{(124)} &\leq \rk(\Um_{2(14)} | \Vm'_2) = \rk([B,B]) - \rk([B]) = 0, \\
        \lambda_{(134)} &\leq \rk(\Um_{1(34)} | \Vm'_1) = \rk([A,A]) - \rk([A]) = 0, \\
        \lambda_{(234)} &\leq \rk(\Um_{2(34)} | \Vm'_2) = \rk([B,B]) - \rk([B]) = 0.
    \end{align*}
    Therefore, $\lambda_{(124)} = \lambda_{(134)} = \lambda_{(234)} = 0$.
    For $\lambda_{(1234)}$, we have
    \begin{align*}
        \lambda_{(1234)} &\leq \rk(\Um_{1(234)}|\Vm'_1) = \rk([A,B]) - \rk([A]) = 1, \\
        \lambda_{(1234)} &\leq \rk(\Um_{2(134)}|\Vm'_2) = \rk([B,C]) - \rk([B]) = 1, \\
        \lambda_{(1234)} &\leq \rk(\Um_{3(124)}|\Vm'_3) = \rk([C,D]) - \rk([C]) = 1, \\
        \lambda_{(1234)} &\leq \rk(\Um_{4(123)}|\Vm'_4) = \rk([A,D]) - \rk([D]) = 1,
    \end{align*}
    thus, $\lambda_{(1234)} = 1$. 
    
    \paragraph*{Stage~3}
    Since all users have been served, $\lambda_k = 0$. That is, $\rk(\Um_{k}|\Vm'_k) = \lambda_{(1234)} \leq 1$ for every $k \in [4]$.
    
    \paragraph*{All together}
    By minimizing the objective function in~\eqref{eq:capacityfor4users}, we attain only $\lambda_{(1234)} = 1$. The server thus sends
    \begin{align*}
        X = (A+B, \ B+C, \ C+D), \text{ with } \qsf=2,
    \end{align*}
    to serve the first three users.
    The sum of all transmissions gives $A+D$ which user~$4$ uses to decode $A$. 
    Thus, we served $4$ users by only $3$ transmissions, 
    saving $1$ transmission compared to serving users individually.

    \section{Conclusion}
    \label{sec:conclusion}
    This paper proposes a new scheme for the LCBC with arbitrary number of users.
    This new scheme is based on a subspace decomposition into linear subspaces. 
    Future work includes  
    comparing our achievable rate with a converse bound;
    further strengthening our scheme by accounting for dependencies that may still exist in the subspaces in our used decomposition thus lowering the rate; and 
    applications to related problems such a coded caching with linear coded placement and with scalar linear function retrieval.

\newpage
\bibliographystyle{ieeetr}
\bibliography{ym-v2}
\appendices

\section{Proof of Lemma~\ref{lem:decompositionK}}
\label{sec:proofdecompositionK}
\begin{figure*}
    \centering
    \begin{subfigure}[b]{0.45\textwidth}
        \centering
        \begin{tikzpicture}
            [level distance=10mm,
            every node/.style={draw=black,rectangle,inner sep=4pt, sibling distance=15mm}
            ]
            \node {$\{1,2,3,4\}$}
            child {node {$\{1,2,3\}$}
                child {node {$\{1,2\}$}
                child {node {$\{1\}$}}
                child {node {$\{2\}$}}
                }
                child {node {$\{3\}$}}
            }
            child {node {$\{4\}$}};
        \end{tikzpicture}
      \caption{\small Binary tree generated for $t=5$ by $\{1,2,3,4\}$.}
      \label{fig:firsttreeK=5}
    \end{subfigure}
    \begin{subfigure}[b]{0.45\textwidth}
        \centering
        \begin{tikzpicture}
            [level distance=10mm,
            every node/.style={draw=black,rectangle,inner sep=4pt, sibling distance=15mm}
            ]
            \node {$\{2,3,4,5\}$}
                child {node {$\{2,3,4\}$}
                    child {node {$\{2,3\}$}
                    child {node {$\{2\}$}}
                    child {node {$\{3\}$}}
                    }
                    child {node {$\{4\}$}}
                }
            child {node {$\{5\}$}};
        \end{tikzpicture}
      \caption{\small Binary tree generated for $t=5$ by $\{2,3,4,5\}$.}
      \label{fig:secondtreeK=5}
    \end{subfigure}
    \caption{\small Generated binary trees reveals dependency relationship among bases.}
\end{figure*}
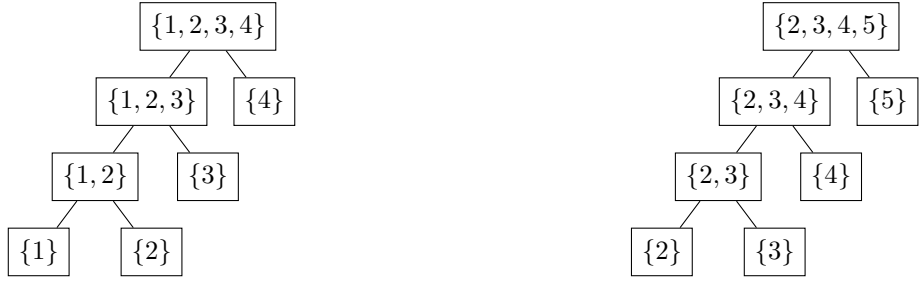

We first redefine the subspace in~\eqref{eq:compositionsubspace}. 
Given a ground set $\Tc$, for any two disjoint subsets $\Sc, \Lc$ from $\Tc$, we define
\begin{align}
\Um_{(\Sc)(\Lc)} = \Um_{(\Sc)} \cap \Um_{(\Lc)},
\end{align}
where $\Um_{(\cdot)}$ is defined in~\eqref{eq:intersectionandunionsubspaces}.
When $\Sc$ or $\Lc$ only contain one element, we omit the parenthesis for the sake of simplicity. 
That is, if $\Sc = \{k\}$, then it is equivalent to $\Um_{k(\Lc)}$.
For example, $\Um_{(1,2)(3,4)}$ denotes a subspace spanned by both $\Um_{(1,2)}$ and $\Um_{(3,4)}$.
Similarly, we extend the definition of basis in~\eqref{eq:basisdefinition} in the same fashion.

We use a graphical method as an aide of proof. Let $t$ be a integer, when $t=3$,
$\{\Bm_{1(2,3)}, \Bm_{2(1,3)}, \Bm_{3(1,2)}\}$ are pairwise linearly independent, that is, any two of them can derive the remaining one, as stated in Lemma~\ref{lem:decompositionK3}.
We can use a binary tree to represent such a linear dependency as shown in~\eqref{eq:lineardependency}.
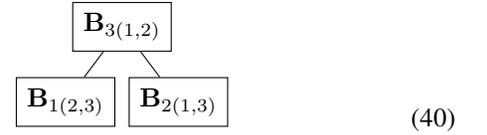
\begin{align}
    \begin{tikzpicture}
        [level distance=10mm,
        every node/.style={draw=black,rectangle,inner sep=4pt, sibling distance=15mm}
        ]
        \node {$\Bm_{3(1,2)}$}
        child {node {$\Bm_{1(2,3)}$}}
        child {node {$\Bm_{2(1,3)}$}};
    \end{tikzpicture}
    \label{eq:lineardependency}
\end{align}
Alternatively, \eqref{eq:lineardependency} states that if we know two leaf nodes, $\Bm_{1(2,3)}$ and $\Bm_{2(1,3)}$, then their parent node $\Bm_{3(1,2)}$ is known. 

For the sake of simplicity, we assign the node $\Sc$, where $\Sc \subseteq [t]$, to %
the basis $\Bm_{\Sc \cap ([t] \setminus \Sc)}$. Both node $\Sc$ or $[t] \setminus \Sc$ represent the same basis.
\eqref{eq:lineardependency2} shows a binary tree which we aim to construct when $t=5$. 
All nodes $\{k\}$, where $k \in [4]$. represents the basis $\Bm_{k([t] \setminus \{k\})}$, and node $\{1,2,3,4\}$ represents the basis $\Bm_{5(1,2,3,4)}$, as defined in Section~\ref{sec:generalsubspacedecomposition}. 
We define the cardinality of a node as the size of the set represented by this given node;
for example, the cardinalities of node $\{1\}$ and $\{1,2\}$ are 1 and 2, respectively.
\begin{align}
    \begin{tikzpicture}
        [level distance=10mm,
        every node/.style={draw=black,rectangle,inner sep=4pt, sibling distance=15mm}
        ]
        \node {$\{1,2\}$}
        child {node {$\{1\}$}}
        child {node {$\{2\}$}};
    \end{tikzpicture}
    \label{eq:lineardependency2}
\end{align}

We connect node $1$ and $2$ to node $\{1,2\}$, since by Lemma~\ref{lem:decompositionK3}, these three bases are pairwise linear independent. Moreover, we always make nodes with cardinality $1$ as leaf nodes, and place nodes with larger cardinality on higher level, as showed in~\eqref{eq:lineardependency2}.
In such a binary tree, if any two children nodes are known, then automatically their parent node is known because of the linear dependence that exists.
For example, in Fig.~\ref{fig:firsttreeK=5}, node $\{1\}$ (representing $\Bm_{1(2,3,4,5)}$) and $\{2\}$ ($\Bm_{2(1,3,4,5)}$) together reveal node $\{1,2\}$ ($\Bm_{(1,2)(3,4,5)}$); node $\{1,2\}$ ($\Bm_{(1,2)(3,4,5)}$) and $\{3\}$ ($\Bm_{3(1,2,4,5)}$) together reveal node $\{1,2,3\}$ ($\Bm_{(1,2,3)(4,5)}$); node $\{1,2,3\}$ ($\Bm_{(1,2,3)(4,5)}$) and $\{4\}$ ($\Bm_{4(1,2,3,5)}$) together reveal node $\{1,2,3,4\}$ ($\Bm_{5(1,2,3,4)}$).

For any arbitrary $t$, if we know all information of the leaf nodes, that is, all bases $\Bm_{k([t]\setminus\{k\})}$ where ${k \in [t]}$, we can derive the root node, representing the basis $\Bm_{t(1,2,\ldots,t-1)}$. Note that we can permute the order of node index, for example, Fig.~\ref{fig:secondtreeK=5} shows another binary tree generated by $t=5$, with different leaf nodes. 
Therefore these $t-1$ bases, represented by leaf nodes, are linearly independent, as claimed in Lemma~\ref{lem:decompositionK}.

\section{Remark on Lemma~\ref{lem:decompositionK}}
\label{sec:remarklemmadecompositionK}

When $\Ksf=3$, the bases in Fig.~\ref{fig:venn1}, where $\{e_1, e_2, e_3\}$ are pairwise independent, can be represented by a convex cone in the space $\RR^7$, whose extreme rays are listed in Table~\ref{tab:convexconeK3}. 
An extreme ray represents a `direction' in the cone that is not reducible to a combination of other directions within the cone.
Each row in Table~\ref{tab:convexconeK3} shows the coordinate of an extreme ray for $\Bm_1, \Bm_2, \Bm_3, \Bm_{12}, \Bm_{13}, \Bm_{23}, \Bm_{123}$ and $(\Bm_{1(23)}, \Bm_{2(13)}, \Bm_{3(12)})$ respectively. Such a convex cone is derived from a representable polymatroid, which is constructed by non-Shannon-type inequalities. It is exactly the entropic region with $3$ random variables in~\cite{hammer2000inequalities}.

Similarly, the subspace decomposition for any arbitrary $\Ksf$ matrices can be represented as a convex cone in the space of $\RR^{2^\Ksf-1}$, where exhaustive search for those extreme rays is possible. \cite[Theorem~5]{hammer2000inequalities} shows the subspace decomposition on any $\Ksf=4$ matrices, and further proves such a convex cone is characterized by the shannon-type inequalities and the Ingleton inequality. %

\begin{table}[h!]
    \centering
    \caption{\small Extreme rays of the convex cone for subspace decomposition with $\Ksf=3$ from~\cite{hammer2000inequalities}.}
    \label{tab:convexconeK3}
    \begin{tabular}{|c|c|c||c|c|c|c|c|c|c|}
    \hline
    \multicolumn{3}{|c|}{\textbf{subspaces}} & \multicolumn{7}{c|}{\textbf{extreme ray}} \\ \hline
    $\Um_1$ & $\Um_2$ & $\Um_3$ & \rotatebox{90}{$\rk(\Um_1)$} & 
    \rotatebox{90}{$\rk(\Um_2)$} & 
    \rotatebox{90}{$\rk(\Um_3)$} & 
    \rotatebox{90}{$\rk([\Um_1, \Um_2])$} & 
    \rotatebox{90}{$\rk([\Um_1, \Um_3])$} & 
    \rotatebox{90}{$\rk([\Um_2, \Um_3])$} & 
    \rotatebox{90}{$\rk([\Um_1, \Um_2, \Um_3])$} \\ \hline
    $\{e_1\}$ & 0 & 0 & 1 & 0 & 0 & 1 & 1 & 0 & 1 \\ \hline
    0 & $\{e_1\}$ & 0 & 0 & 1 & 0 & 1 & 0 & 1 & 1 \\ \hline
    0 & 0 & $\{e_1\}$ & 0 & 0 & 1 & 0 & 1 & 1 & 1 \\ \hline
    $\{e_1\}$ & $\{e_1\}$ & 0 & 1 & 1 & 0 & 1 & 1 & 1 & 1 \\ \hline
    $\{e_1\}$ & 0 & $\{e_1\}$ & 1 & 0 & 1 & 1 & 1 & 1 & 1 \\ \hline
    0 & $\{e_1\}$ & $\{e_1\}$ & 0 & 1 & 1 & 1 & 1 & 1 & 1 \\ \hline
    $\{e_1\}$ & $\{e_1\}$ & $\{e_1\}$ & 1 & 1 & 1 & 1 & 1 & 1 & 1 \\ \hline
    $\{e_1\}$ & $\{e_2\}$ & $\{e_3\}$ & 1 & 1 & 1 & 2 & 2 & 2 & 2 \\ \hline
    \end{tabular}
\end{table}

\section{Proof of Lemma~\ref{lem:multicastmessage}}
\label{sec:proofmulticastmessage}

\begin{lem} \rm \label{lem:multicastmessage}
Consider full rank where all matrices $\Am \in \FF_\qsf^{d \times a}$ and $\{\Bm_k \in \FF_\qsf^{d \times b_k}\}_{k \in [t]},$ that is, $\rk[\Am]=a,$ and $\rk[\Bm_k]=b_k, \ \forall k \in [t]$.
There exist submatries $\{\Bm'_k\}_{k \in [t]}$ where $\Bm'_k \subseteq \Bm_k$, $\Bm'_k \in \FF_\qsf^{d \times n_k}$, and $\rk(\Bm'_k) = n_k$, such that $[\Am, \Bm'_1, \ldots, \Bm'_t]$ has full rank $a+n_1 + \ldots + n_t$, if the non-negative integers $n_1, \ldots, n_t$ satisfy %
$\sum_{k \in \Lc} n_k \leq \rk(\Bm_{\ell_1}, \ldots, \Bm_{\ell_j} | \Am)$ for every $\Lc = \{\ell_1,\ldots, \ell_j\}\subseteq [t]$.
\end{lem}

We prove Lemma~\ref{lem:multicastmessage} by induction. 

\paragraph{Step~1}
When $t=1$, our lemma states that 
we can find a submatrix of $\Bm_1$, denoted by $\Bm'_1 \in \FF_\qsf^{d \times n_1}$  and with $\rk[\Bm'_1] = n_1$, such that $[\Am, \Bm'_1]$ is full rank, 
that is, $\rk[\Am, \Bm'_1] = a + n_1$, if $n_1 \leq \rk(\Bm_1 | \Am)$. This claim was proved in~\cite[Corollary 2]{yao2024capacity} as a corollary to~\cite[Lemma~4]{yao2024capacity}, which is our lemma for $t=2$. The proof of~\cite[Lemma~4]{yao2024capacity} leveraged the Steinitz Exchange lemma.

\paragraph{Step~2}
We assume we have found the first $p-1$ eligible submatrices
$\Bm'_k \subseteq \Bm_k : \rk(\Bm'_k) = n_k$ for $k\in[p-1]$
such that $[\Am, \Bm'_1, \ldots, \Bm'_{p-1}]$ is a full rank matrix
where the non-negative integers $n_1, \ldots, n_{p-1}$ satisfy 
$\sum_{k \in \Lc} n_k 
\leq \rk(\Bm_{\ell_1}, \ldots, \Bm_{\ell_j} | \Am)$ 
for every $\Lc = \{\ell_1,\ldots, \ell_j\}\subseteq [p-1]$.

\paragraph{Step~3}
Next we aim to find the $p$-th submatrix 
$\Bm'_p \subseteq \Bm_p : \rk(\Bm'_p) = n_p$
such that $[\Am, \Bm'_1, \ldots, \Bm'_{p}]$ is a full rank matrix
with $\rk(\Bm'_{\ell_1}, \ldots, \Bm'_{\ell_j} | \Am) 
= \sum_{k \in \Lc} n_k 
\leq \rk(\Bm_{\ell_1}, \ldots, \Bm_{\ell_j} | \Am)$ 
for every $\Lc = \{\ell_1,\ldots, \ell_j\}\subseteq [p]$.

For every $\Lc = [\ell_1, \ldots \ell_j] \subseteq [p-1]$
\begin{align}
    n_{p} 
    &\leq \rk(\Bm_{p} | \Am, \Bm'_{\ell_1}, \ldots, \Bm'_{\ell_j})  \\
    &= \rk(\Bm_{p}, \Bm'_{\ell_1}, \ldots, \Bm'_{\ell_j} | \Am) - \rk(\Bm'_{\ell_1}, \ldots, \Bm'_{\ell_j} | \Am) \\
    &= \rk(\Bm_{p}, \Bm'_{\ell_1}, \ldots, \Bm'_{\ell_j} | \Am) - \sum_{k \in \Lc} n_k  \\
    &\leq \rk(\Bm_{p}, \Bm_{\ell_1}, \ldots, \Bm_{\ell_j} | \Am) - \sum_{k \in \Lc} n_k.
\end{align}

We continue this process until $p=t$ as claimed in Lemma~\ref{lem:multicastmessage}. 
Note that Lemma~\ref{lem:multicastmessage} implies $\rk(\Bm'_{\ell_1}, \ldots, \Bm'_{\ell_j} | \Am) = 
\sum_{k \in \Lc} n_k$ for every $\Lc \subseteq [t]$, as $[\Am, \Bm'_1, \ldots, \Bm'_t]$ is a full-rank matrix.

\section{Proof of Theorem~\ref{thm:achievabilityLCBCK}}
\label{sec:proofachievabilityLCBCK}

Recall that the dimension of a single data stream is $\dsf$. 
We focus on user~1 and explain how to derive the constraints~\eqref{eq:firsttypeconstraint}, \eqref{eq:secondtypeconstraints} and~\eqref{eq:lasttypeconstraint}.

\paragraph*{Stage~1} 
By Lemma~\ref{lem:multicastmessage}, we can find $\Um'_{[\Ksf]} \in \FF_\qsf^{\dsf \times \lambda_{[\Ksf]}}$ with full rank, which is a submatrix of $\Um_{[\Ksf]}$, such that,
\begin{align}
    \rk([\Vm_1', \Um'_{[\Ksf]}]) = m'_k + \lambda_{[\Ksf]} \leq m'_k + \rk(\Um_{[\Ksf]} | \Vm'_1). \label{eq:firstmulticast}
\end{align}
We continue to the next subspace. For every $\Tc \in \Omega_{[\Ksf]}^{\Ksf-1}$ and $1 \in \Tc$, we can find  $\Um'_{\Tc} \in \FF_\qsf^{\dsf \times \lambda_{\Tc}}$, which is a submatrix of $\Um_{\Tc}$ with full rank,  such that,
\begin{multline}
    \rk([\Vm_1', \Um'_{[\Ksf]}, \Um'_{\Tc}]) = m'_k + \lambda_{[\Ksf]} + \lambda_{\Tc} \\ \leq m'_k + 
     \rk(\Um_{\Tc} | \Vm'_1). \label{eq:secondmulticast}
\end{multline}
Note that $\mathsf{LC}(\Um_\Tc) = \{\Um_{[\Ksf]}\}$, then~\eqref{eq:firstmulticast} and~\eqref{eq:secondmulticast} correspond to the constraints~\eqref{eq:firsttypeconstraintmain} and~\eqref{eq:firsttypeconstraintsubset}, respectively, induced by $\Um_\Tc \in \Vc_1$.
That is, let $t = \Ksf$ first, for every $\{\Um_{\Sc}: \Sc \in \Omega_{[\Ksf]}^{t}, 1 \in \Sc\} \subseteq \Vc_1$,
this process starts from $\Um_{[\Ksf]}$ (that is, $t=\Ksf$), keeps iterating with $t$ decreasing, until $t=2$;
by doing so we attain the constraints~\eqref{eq:firsttypeconstraintmain} and~\eqref{eq:firsttypeconstraintsubset} with $k=1$ respectively. 

\paragraph*{Stage~2} Similarly, let $t=3$ first, for every $\{\Um_{1(\Tc)}: \Tc \in \Omega_{[\Ksf]}^{t-1}\}$, says $\Um_{1(23)}$, by Lemma~\ref{lem:multicastmessage}, we can find %
$\Um'_{1(23)} \in \FF_\qsf^{\dsf \times \lambda_{(123)}}$, which is a submatrix of $\Um_{1(23)}$ with full rank, such that
\begin{align}
    &\rk([\Vm'_1, \Um'_{[\Ksf]}, \ldots, \Um'_{12}, \Um'_{13}, \Um'_{1(23)}]) \notag 
    \\ &= m'_k + \lambda_{(123)} +\sum_{\Tc \subseteq [\Ksf], |\Tc| \geq 2, \{1,2\} \subseteq \Tc, \{1,3\} \subseteq \Tc} \lambda_{\Tc} \notag
    \\ &= m'_k + \text{"sum of $\mathsf{C}_\lambda(\Um_{1(23)})$"} \notag \\ 
    &\leq m'_k + \rk(\Um_{1(23)} | \Vm'_1), 
    \label{eq:firstcomplexmulticastmain} 
\end{align}
and
\begin{align}
    &\rk([\Vm'_1, \Um'_{[\Ksf]}, \ldots, \Um'_{12}, \Um'_{13}]) \notag 
    \\ &= m'_k +\sum_{\Tc \subseteq [\Ksf], |\Tc| \geq 2, \{1,2\} \subseteq \Tc, \{1,3\} \subseteq \Tc} \lambda_{\Tc} \notag
    \\ &= m'_k + \text{"sum of $\mathsf{C}_\lambda(\Um_{12}) \cup \mathsf{C}_\lambda(\Um_{13})$"} \notag \\ 
    &\leq m'_k + \rk(\Um_{12}, \Um_{13} | \Vm'_1), 
    \label{eq:firstcomplexmulticastsubset}
\end{align}
where~\eqref{eq:firstcomplexmulticastmain} and~\eqref{eq:firstcomplexmulticastsubset} are  equivalent to~\eqref{eq:secondtypeconstraintsmain} and~\eqref{eq:secondtypeconstraintssubset}, respectively, induced by $\Um_{1(23)}$. 
We proceeds similarly with $t$ increasing until $t=\Ksf$. 
By doing so we attain the constraints~\eqref{eq:secondtypeconstraintsmain} and~\eqref{eq:secondtypeconstraintssubset} with $k=1$.

\paragraph*{Stage~3} 
In the end, some data are demanded by each user itself, multicast is not possible and it must be transmitted individually. Therefore, by Lemma~\ref{lem:multicastmessage}, we can find a submatrix of $\Um_1$, namely $\Um'_1 \in \FF_\qsf^{\dsf \times \lambda_1}$, such that
\begin{align}
    &\rk([\Vm'_1, \Um'_{[\Ksf]}, \ldots, \Um'_{1(23)}, \ldots, \Um'_1 ]) \notag 
    \\ &= m'_k +  \sum_{\Sc \in [\Ksf], 1 \in \Sc} \lambda_\Sc + \sum_{\Sc \subseteq [\Ksf], |\Sc| \geq 3, 1 \in \Sc} \lambda_{(\Sc)} \notag
    \\ &= m'_k + \text{"sum of $\mathsf{C}_\lambda(\Um_{1})$"} \notag \\ 
    &= m'_k + \rk(\Um_{1} | \Vm'_1), \label{eq:lastunicast}
\end{align}
where~\eqref{eq:lastunicast} is equivalent to~\eqref{eq:lasttypeconstraint} induced by $k=1$.

We repeat all stages over for every user $k \in [\Ksf]$, thus we obtain the constraints in~\eqref{eq:firsttypeconstraint}, \eqref{eq:secondtypeconstraints} and~\eqref{eq:lasttypeconstraint}.

\paragraph*{Construction of Multicast Messages} 
For every $V = \Um_\Sc \in \Vc_1$, we let $\Mm_{\Sc} \in \FF_\qsf^{\rk(\Um_\Sc) \times \lambda_{\Sc}}$, so
    $\Um'_{\Sc} = \Um_{\Sc} \Mm_{\Sc}$
is a submatrix which constructs a multicast message to serve the users in~$\Sc$.
For every $V = \Um_{k(\Sc)} \in \Vc_2$, %
by the general subspace decomposition in Section~\ref{sec:generalsubspacedecomposition}, we know
\begin{align*}
    \langle \Um_{k(\Sc)} \rangle &= \langle \mathsf{C}(\Um_{k(\Sc)}) \rangle \\
    &= \big[\Um_{kj}: j \in \Sc \big] \cup \big[\Bm_{k(\Pc)}: \Pc \subseteq \Sc, |\Pc| \geq 2 \big].
\end{align*}
Thus, $\Um'_{k(\Sc)}$ can be rewritten as,
\begin{align}
    \sum_{j \in \Sc} \alpha_{kj}^{(\Sc)} \Nm_{kj} \Um_{kj}  + \sum_{\Pc \subseteq \Bc, |\Pc| \geq 2} \alpha_{k,\Pc}^{(\Sc)} \Mm_{\{k\}\cup\Pc} \Bm_{k(\Pc)},
\end{align}
where $\alpha_{\star}$ are coefficients either $-1$ or $+1$, and
\begin{align}
    \Nm_{kj} &\in \FF_\qsf^{\rk(\Um_{kj}) \times \lambda_{kj}}, \ \forall j \in \Sc, \\
    \Mm_{k(\Pc)} &\in \FF_\qsf^{\rk(\Bm_{k(\Pc)}) \times \lambda_{\{k\}\cup\Pc}}, \ \forall \Pc \subseteq \Sc, |\Pc| \geq 2. 
\end{align}
In general, for any $\Tc \subseteq [\Ksf]$ and $|\Tc| := t \geq 3$, we can always find submatries for users~$\Tc$, such that
\begin{align}
    \sum_{k \in \Tc} \Um'_{k(\Tc\setminus\{k\})} = 0. \label{eq:sumofcomplexmulticastmessages}
\end{align}
More specifically, in the LHS of~\eqref{eq:sumofcomplexmulticastmessages}, we notice that
\begin{itemize}
    \item $\Nm_{ij} \Um_{ij}$ appears twice for every $\{i,j\} \subseteq \Tc$;
    \item for every $\Pc \subseteq \Tc$ and $|\Pc| \geq 3$, $\{\Mm_{\Pc} \Bm_{k(\Pc \setminus \{k\})}: k \in \Pc\}$ appear only once.
\end{itemize}
Therefore, by Lemma~\ref{lem:decompositionK} and alternating those sign coefficients $\alpha_\star$ properly, \eqref{eq:sumofcomplexmulticastmessages} holds true.

For every $V = \Um_k \in \Vc_3$, we let $\Um'_k = \Um_k\Mm_k$ be the submatrix which constructs unicast message for user $k$ only, where $\Mm_k \in \FF_\qsf^{\rk(\Um_k) \times \lambda_k}$. $\Um'_k$ contains information of $\Vm_k$ which does not deduce from the previous multicast messages.

Thus, the transmit signal  $X$ is composed of multicast messages as follows
\begin{multline*}
    X = \{\Fm \Um'_{\Sc}: \Sc \subseteq [\Ksf]\} \\ 
    \cup \left(\bigcup_{t \geq 3} \text{"any $t-1$ of } 
    \{\Fm \Um'_{k(\Sc \setminus \{k\})}: \Sc \in \Omega_{[\Ksf]}^{t}\}\text{"}\right),
\end{multline*} 
and the achievable rate is
as shown in~\eqref{eq:objectivefunctionloadsum}.

\paragraph*{Decoding Correctness}
For the case of $\Ksf=3$ users, \cite[Section 8.5]{yao2024capacity} put forth a random-coding-like argument that shows that ``mixing of information is possible to reconcile the users' different perspectives.'' This is so because the argument focuses on one user, say user~1, and find submatrices that work for said user;  however these choices may not work for the remaining users. This challenge can be overcome by choosing symbols in the extension field to jointly code over a sufficiently large number of computations.

Let $P_1$ be a polynomial with elements of all $\Nm_\star$ and $\Mm_\star$ as variables,
\begin{align}
    P_1 = \det\big[\Vm'_1, \Um'_{[\Ksf]}, \ldots, \Um'_{1(23)}, \ldots, \Um'_1, \Zm_1\big],
\end{align}
where $\Zm_1 \in \FF_\qsf^{(\dsf - m_1 - m'_1) \times \dsf}$ is a submatrix of $\Id^{\dsf \times \dsf}$.
Similar for $P_2, \ldots, P_\Ksf$. Let $P = \prod_{k\in[\Ksf]} P_k$, the degree of $P$ at most $\Ksf\dsf$.
If all $\Mm_\star$ and $\Nm_\star$ are chosen i.i.d uniformly from $\FF_{\qsf^z}$, where $z \geq \log_\qsf(\Ksf \dsf)$, 
the probability of $P=0$ vanishes.
Thus each polynomial $P_k$, where $k \in [\Ksf]$, is non-zero for appropriate choices of the variables $\Mm_\star$ and $\Nm_\star$.
With this one concludes that all users can restore their desired information. 
The same claim holds true for our construction.

\end{document}